\documentclass{aa}  

\usepackage{graphicx}
\usepackage{hyperref}
\usepackage{txfonts}
\usepackage{booktabs}
\usepackage{amsmath}
\usepackage{amssymb}
\usepackage{textcomp}
\usepackage{xcolor}

\begin{document} 
\title{Unveiling the kinematics of a central region in the triple AGN host NGC 7733-7734 interacting group}

   \author{Saili Keshri
          \inst{1}\fnmsep\inst{2} 
          \and
          Sudhanshu Barway
          \inst{1}
          \and
          Mousumi Das
          \inst{1}
          \and
          Jyoti Yadav
          \inst{1}\fnmsep\inst{2}
          \and
          Francoise Combes
          \inst{3}
          }

   \institute{Indian Institute of Astrophysics, Kormangala II Block, Bengaluru, India, 560034\\
            \email{saili.keshri@iiap.res.in}
         \and
             Department of Physics, Pondicherry University, R.V.Nagar, Kalapet, Puducherry, India, 605014
        \and
            Observatoire de Paris, LERMA, College de France, CNRS, PSL University, Sorbonne University, F-75014 Paris, France\\
             }

\abstract {We present a detailed study of the interacting triple active galactic nuclear system NGC 7733-34, focusing on stellar kinematics, ionised gas characteristics and star formation within the central region and stellar bars of both galaxies. We performed a comprehensive analysis using archival data from MUSE, HST/ACS, and DECaLS, complemented by observations from UVIT and IRSF. We identified a disc-like bulge in both NGC 7733 and NGC 7734 through 2-D decomposition. A central nuclear structure, with a semi-major axis of $\sim$1.113 kpc, was detected in NGC 7733 via photometric and kinematic analysis, confirmed by the strong anti-correlation between $V/\sigma$ and $h_{3}$, indicative of circular orbits in the centre. NGC 7734 lacks a distinct nuclear structure. The presence of disc-like bulge results in an anti-correlation between $V/\sigma$ and $h_{3}$ along with diffuse light. However, it does show higher central velocity dispersion, possibly attributed to an interaction with a smaller clump, which is likely a fourth galaxy within the system. Both galaxies demonstrate ongoing star formation, evidenced by $FUV$ and $H\alpha$ observations. NGC 7734 shows recent star formation along its bar, while NGC 7733 experiences bar quenching. The star formation rate (SFR) analysis of NGC 7734 reveals that the bar region's SFR dominates the galaxy's overall SFR. Conversely, in NGC 7733, the lack of star formation along the bar and the presence of a Seyfert 2 active galactic nuclei at the galaxy centre leave the possibility of a connection between both facts. However, it does not affect the galaxy's overall star formation. Our findings provide valuable insights into the stellar and gas kinematics, star formation processes, and active galactic nuclear feedback mechanisms in interacting galaxies hosting stellar bars.}

\keywords{galaxies: interaction - galaxies: active - galaxies:  interacting - galaxies: individual: NGC 7733 – galaxies: individual: NGC 7734 - techniques: imaging spectroscopy}
    
\maketitle
\section{Introduction}\label{sec:Sec_1}
In our nearby Universe, galaxy interactions have been observed to significantly impact galaxy evolution. When galaxies come close, the gravitational force causes distortions and can lead to the exchange of dust and gas. The gravitational forces from these encounters generate tidal forces, which in turn trigger the collapse and compression of molecular gas clouds. This process is crucial for forming new stars \citep{Mihos96, Barnes96}. Close interactions, particularly involving galaxies with substantial reserves of cold gas, can lead to increased star formation (hereafter - SF) and active galactic nuclear (hereafter - AGN) activity. Such interactions result in enhanced SF compared to isolated galaxies \citep{Kennicutt08, Ellison08, Scudder12}. Additionally, the merger of galactic discs can create channels for gas inflows towards central regions, fueling starbursts \citep{DiMatteo07, Hopkins08}. Studies indicate that the star formation rate (hereafter - SFR) per unit mass in interacting galaxies is approximately 10 times higher than that of isolated systems \citep{Young93}.

Galactic structures such as bulges, bars, discs and spiral arms can be strongly affected by interactions, which in turn affect the distribution of molecular gas and SF, although their effect on star formation efficiency is found to be small \citep{Querejeta21}. Nearly 70\% of disc galaxies in the nearby universe host bars in their central region which is evident in near-infrared images \citep{deVaucouleurs63, Knapen00, Eskridge00, Men07, Nair10, Masters11}, and bars play a vital role in galaxy evolution either by enhancing or suppressing the SF in their host galaxy. There is no uniform picture of where the SF occurs in galactic bars. It can occur in narrow lanes or have a curved morphology, either along one edge or in the middle of the bar \citep{Diaz20}. Bars have statistically been linked to enhanced gas concentration and very clearly linked to enhanced SF in the central kpc region \citep{Jogee05, Lin17}. Studies by \citet{Elmegreen85, Men07} suggest that the dependence of SFR enhancement on morphological type may be due to the longer and stronger (more elongated) bars. Simulations also indicate that only strong bars are effective at funnelling gas to the inner kpc \citep{Regan04} and triggering SF. \citet{Fraser-McKelvie20} have shown the presence of $H\alpha$ (an indicator of massive SF) along the bar length and at the bar ends in a sample of barred galaxies from SDSS-IV MaNGA. \citet{Diaz20} found SF within bars and concluded that star-forming bars are ubiquitous among late-type gas-rich galaxies. \citet{Barway20} found that bars in S0 galaxies appear bluer than those in spiral galaxies, likely due to their environments. In their study most blue barred S0 galaxies are in intermediate-density regions, where minor mergers and tidal interactions are more frequent. These external effects may trigger SF in these bars, hence the bars appear bluer. Thus, interacting galaxy systems present an ideal framework for investigating the potential impact of interactions on the various properties of individual galaxies within the system and their evolutionary processes. 

NGC 7733-34 is a notable example of a pair of barred interacting galaxies and has been identified as a triple AGN system by \citet{Yadav21}. The system comprises the barred spiral galaxies NGC 7733 and NGC 7734. NGC 7733 is situated at a distance of 152 Mpc and the distance of NGC 7734 is 158 Mpc, with NGC 7734 being approximately 48.3 kpc from NGC 7733. Both galaxies have undergone a recent SF burst in their inner discs, and have extended rings and tidal arms. Additionally, NGC 7733 is interacting with a smaller galaxy, NGC 7733N, which appears as an early-type galaxy. Emission line analysis from the central regions of NGC 7733 and NGC 7734 indicates the presence of Seyfert and low-ionisation nuclear emission-line regions (LINER) type AGN activity. NGC 7733N also exhibits Seyfert-like emission characteristics. In a study of southern interacting galaxies, \citet{Horellou97}  estimated the atomic gas (HI) content of the NGC 7733-34 system to be less than $4.2 \times 10^{9} M_{\odot}$. In contrast, the molecular gas content, specifically $^{12}CO(1-0)$, was found to be less than $1.54 \times 10^{9} M_{\odot}$ in NGC 7733 and approximately $1.073 \times 10^{10} M_{\odot}$ in NGC 7734. A photometric study by \citet{Jahan-Miri01}, utilizing UBVRI photometry, revealed that the stellar population in NGC 7733 is significantly younger than that in NGC 7734. 

\citet{Yadav23} conducted a far-ultraviolet ($FUV$) and optical study of the star-forming complexes within this interacting galaxy system. They reported the mean $FUV$ and $H\alpha$ star formation rate surface density (hereafter - $\Sigma_{SFR}$) for NGC 7733 to be 0.008 and 0.194 M$_\odot$yr$^{-1}$kpc$^{-2}$, respectively, and for NGC 7734 to be 0.014 and 0.200 M$_\odot$yr$^{-1}$kpc$^{-2}$. In the present study, we investigate the photometric properties of the individual galaxies in the interacting group NGC 7733-34 to understand the structural properties of the central region, including bulge and stellar bar. We also present a kinematic study of this group to shed light on the system's stellar and ionised gas properties. Our analysis utilizes $FUV$ observations from the Ultraviolet Imaging Telescope (UVIT), archival optical data obtained with the Very Large Telescope's (VLT) Multi-Unit Spectroscopic Explorer (MUSE) integral field spectrograph (IFS), archival optical imaging data from Dark Energy Camera Legacy Survey (DECaLS) and Hubble Space Telescope (HST), and near-infrared (NIR) data from the Infrared Survey Facility (IRSF) telescope. 

This paper is organised as follows: Section~\ref{sec:Sec_2} briefly overviews the data, and Section~\ref{sec:Sec_3} presents the analysis performed for this work. Section~\ref{sec:Sec_4} focuses on the results and discussion, and a summary of the results is presented in Section~\ref{sec:Sec_5}. An appendix includes SF in different regions of NGC 7734. Throughout this work, we used a flat $\Lambda$CDM cosmology with $\Omega_M$ = $0.3$, $\Omega_{\lambda}$ = $0.7$ and Hubble constant (H$_0$) = $70km$$s^{-1}$$Mpc^{-1}$.


\section{Data} \label{sec:Sec_2}
\begin{figure}
    \includegraphics[width=\columnwidth]{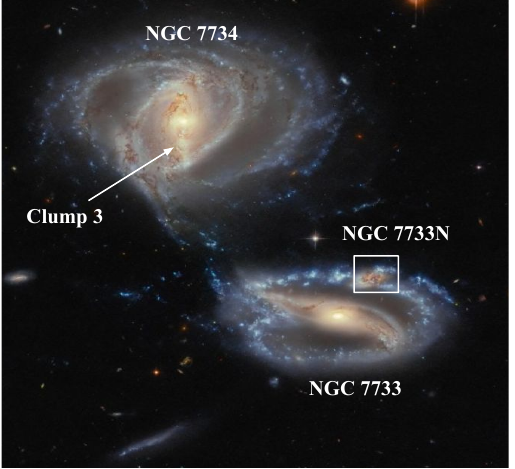}
    \caption{Colour image of the NGC 7733-34 galaxy group taken by NASA's Hubble Space Telescope. The three galaxies NGC 7733, NGC 7733N, and NGC 7734 are marked in this image. Also, the clump present along the bar of NGC 7734 is marked here and has been discussed in Sec.~\ref{sec:Sec_4.3}.}
    \label{fig:HST}
\end{figure}

In this study we have used $FUV$ data obtained from UVIT onboard AstroSat \citep{Agrawal06}, the imaging optical data from Data Release 10 (DR10) of DECaLS \citep{Dey19} and the optical IFU data from MUSE \citep{Bacon06}. The galaxy group was observed on 2019-09-08, and the program ID is 0103.A-0637. We have used the science-ready MUSE-DEEP data cube of the galaxy group from the ESO archive science portal\footnote{https://archive.eso.org/scienceportal/home}. We have also used recently published high-resolution optical images from the Advanced Camera for Surveys (ACS) onboard HST. It is observed with filter $F606W$, corresponding to a broad $V$ band. In Fig.~\ref{fig:HST}, we show the colour image of the galaxy group NGC 7733-34 taken from NASA Science\footnote{https://science.nasa.gov/missions/hubble/hubble-captures-a-galactic-dance-2/}. Three galaxies NGC 7733, NGC 7733N and NGC 7734 are marked here. The NIR imaging data is observed from the IRSF telescope at the South African Astronomical Observatory (SAAO) \citep{Nagayama03}. The detailed discussion on multi-wavelength data used for this work can be found in \citet{Yadav23}. 

\begin{figure*}
    \includegraphics[width=\textwidth]{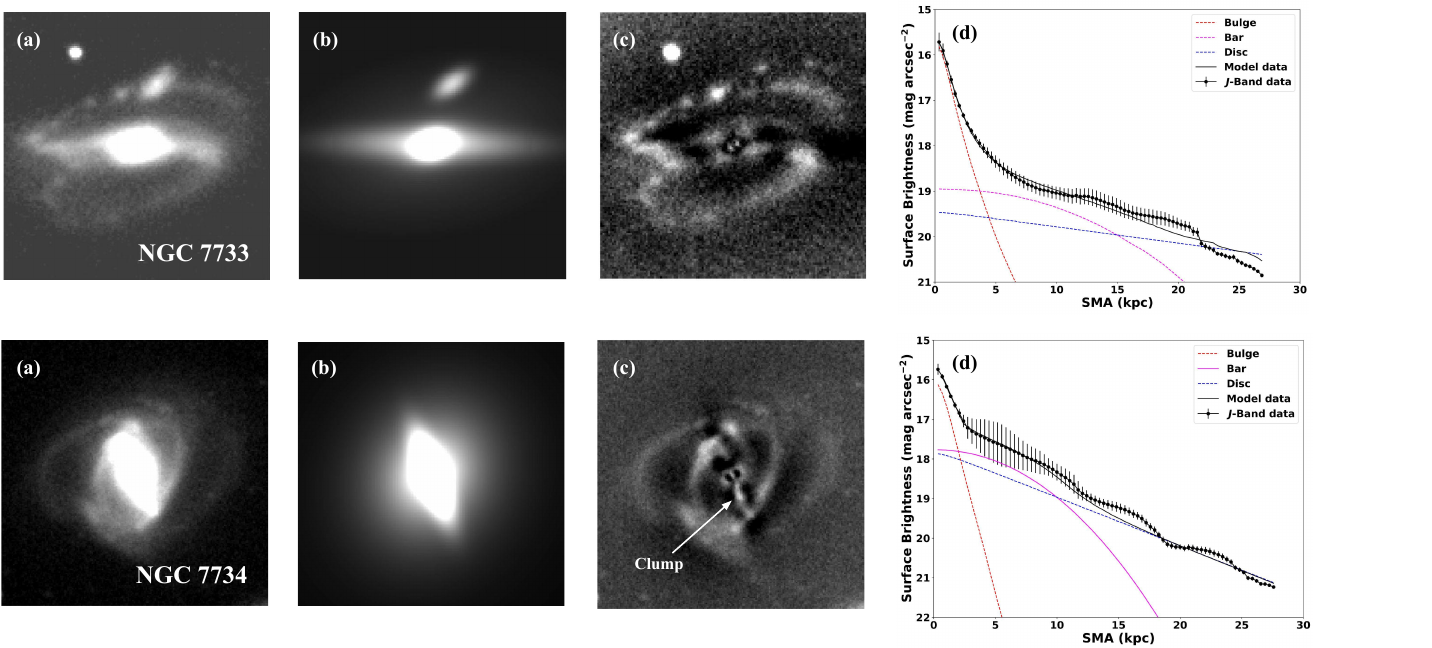}
    \caption{The output of the GALFIT model along with the azimuthally-averaged surface brightness profile of NGC 7733 (top panel) and NGC 7734 (bottom panel) is shown. 
    In both the top and bottom panels, figure (a) is the $J-$ band image of the galaxies, figure (b) is the model image and (c) represents the residual obtained from the GALFIT modelling. The parameters of the models of NGC 7733 and NGC 7734 are tabulated in Table~\ref{tab:Table_1}. The results of ellipse fitting are presented in figure (d). Azimuthally averaged surface brightness of the best-fitted ellipse plotted as a function of the semi-major axis of the ellipse (solid dots), here the error bars represent the rms error of the intensity measured along each fitted isophote. The surface brightness profile is decomposed into two Sersic and one exponential component (in both the panels), a bulge (red dashed), a bar (magenta dashed) and a disc (blue dashed). The sum of these three components is shown by a black solid line.  The clump is marked in the residual image in figure (c) of the bottom panel. North is up, and
    east is to the left.}
    \label{fig:galfit}
\end{figure*}


\section{Analysis}\label{sec:Sec_3}
In this section, we shall determine the properties of the structural components (bulge, bar and disc) of NGC 7733 and NGC 7734. These parameters were obtained from $J$- and $K_{s}$ band NIR images of the galaxies. We discuss the presence of nuclear structures obtained using HST/$F606w$ band images. We present a brief discussion on the analysis of the MUSE data cube. Additionally, we will discuss the detection of star-forming clumps (hereafter - SFCs) using $FUV$ and $H\alpha$ images of NGC 7733 and NGC 7734.

\subsection{Isophotal Analysis}\label{sec:Sec_3.1}
Stellar bars in galaxies are mainly characterised by their length, strength and pattern speed. Besides the bar length and pattern speed, the bar strength is a crucial characteristic. It signifies the bar's contribution to the overall potential of the galaxy, which is hard to measure. But, there are many proxies to measure bar strength. For example, the maximum ellipticity \citep{Laurikainen02, Erwin04} and the boxiness of the bar isophotes \citep{Gadotti11} have been used to estimate bar strength. Bar strength can also be estimated by measuring the torque exerted by the bar \citep{Combes81, Buta01, Laurikainen02} or by measuring the amplitude of the m=2 Fourier mode \citep{Athanassoula03}. In this work, we are using one of the simplest definitions as given by \citet{Abraham00} and later used by \citet{Aguerri09}, as it is easy to determine from the ellipse fits. This method requires the measurement of the bar ellipticity ($\epsilon_{bar}$) at $r_{bar}$.

In order to measure the bar strength, first we need to deproject the galaxies according to the galaxy inclination (i). To estimate i of NGC 7733 and NGC 7734, we performed the isophotal analysis on $J$-band image (since, the $J$-band NIR images of NGC 7733 and NGC 7734 a have high signal-to-noise ratio (S/N) compared to other NIR images, a typical feature of IRSF observations (see, \citet{Nagayama03}). The ellipse-fitting began a few arcsec away from the centre and extended to the outermost isophotes of the disc. All parameters, including the centres, were allowed to vary during the fitting. It allowed us to establish the ellipticity ($\epsilon$), position angle (PA), and mean intensity of the best-fit ellipse as a function of the semi-major axis. We estimated the $\epsilon$ corresponding to a surface brightness of 21 $mag/arcsec^{2}$, which is consistent with the surface brightness of 25 $mag/arcsec^{2}$ in B-band \citep{Jarrett03}.

We determined, i, using $cos \ i = (1-\epsilon$). Now, we deproject the $J$-band NIR image of NGC 7733 into the sky plane using i = 57.23\textdegree and PA = 103.9\textdegree. Similarly, for NGC 7734, the deprojection has been done using i = 32.14\textdegree, PA = 144\textdegree. Once we deproject galaxy images, we use them for further analysis.

Several methods have been developed to measure the bar's length. The three most popular methods are - the Fourier decomposition, the PA radial profile and the ellipticity radial profile. To estimate the bar length using Fourier decomposition, the Fourier analysis of azimuthal luminosity profiles is used \citep{Ohta90, Aguerri00}. The other two methods use the isophotal analysis technique of fitting ellipses to photometric isophotes. In the PA radial profile method, the isophotes associated with the bar show nearly constant PA and then it changes to the outer disc orientation at large radii \citep{Wozniak95, Aguerri00A}. The bar length is determined at the radius where the PA changes by $\Delta$PA from the value corresponding to the local maxima of $\epsilon$. The value of $\Delta$PA used in literature is 5\textdegree \citep{Aguerri15, Guo19}. The radial ellipticity profile of a barred galaxy exhibits a rise from a central value of zero to a local maximum, where the bar dominates the isophotes. Then, it transitions to the disc-dominated isophotes and decreases to a local minimum, which corresponds to the end of the bar. Thus the local maximum represents the lower limit and the local minimum represents the upper limit of the bar length \citep{Michel-Dansac6}. In this study, We adopt the radius reaching the local maximum $\epsilon$ where the PA of the bar isophotes' remains constant \citep{Barazza08, Aguerri09}, as the bar length and the corresponding $\epsilon$ as bar ellipticity ($\epsilon_{bar}$). The bar strength ($f_{bar}$) was obtained using the bar's axis ratio (b/a: where a and b are the semi-major and semi-minor axis, respectively), in line with the relation outlined in \citet{Aguerri09} which is derived from the maximum value of the ellipticity profile along the bar, which usually happens at the end of the bar. The value of bar strength ranges from zero for an unbarred galaxy to close to unity for a strong bar \citep{Abraham00}. 

\begin{equation}
    f_{bar} = \frac{2}{\pi} [arctan\; (1-\epsilon_{bar})^{-1/2}\, - \, arctan\; (1-\epsilon_{bar})^{+1/2}]
    \label{eq:1}
\end{equation}

The $\epsilon_{bar}$ ($\epsilon_{bar}$ = 1 - $(b/a)_{bar}$) represents the maximum value of the ellipticity profile along the bar. Studies \citep{Laurikainen02A, Laurikainen07} have shown that $\epsilon_{bar}$ correlates quite well with the measurement based on the maximum relative tangential force due to the bar \citep{Buta01}.

We have determined the bar strength ($f_{\text{bar}}$) to be 0.538 $\pm$ 0.007 for NGC 7733 and 0.267 $\pm$ 0.005 for NGC 7734. Since NGC 7733 is a highly inclined galaxy with a close companion, the bar ends are being affected due to the ongoing interaction with NGC 7734 as visible in Fig.~\ref{fig:HST}. Hence the bar ends seem to be distorted and bar length and strength measurements are done as best as possible. The bar length and strength observed in the NGC 7733-34 system are consistent with the characteristics of late-type galaxies reported by \citet{Diaz16}, based on 3.6-$\mu$m imaging from the Spitzer Survey of Stellar Structure in Galaxies ($S^{4}G$). Notably, NGC 7733 displays a stronger bar compared to NGC 7734. This observation is particularly intriguing given that Seyfert galaxies generally do not exhibit strong bars \citep{Shlosman00}. This discrepancy invites further numerical and theoretical investigations to explore the relationship between nuclear activity and stellar bar evolution. 

\begin{figure}
    \hspace{1.5cm}
    \includegraphics[width=6cm,height=6cm]{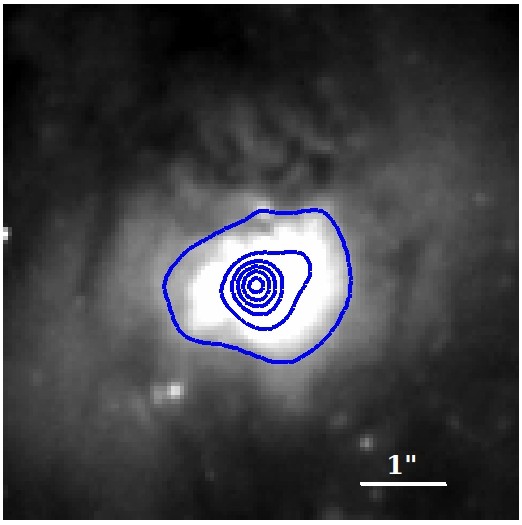}
    \caption{The $F606W$ filter image of the central 6 arcsec $\times$ 6 arcsec region of NGC 7734. The contour is overlaid in the image to show the presence of central very central circular contours. At the distance of NGC 7734, 1 arcsec = 766 pc. North is up, and east is to the left.}
    \label{fig:HST_Centre}
\end{figure}
\begin{figure*}
    \hspace{3.0 cm}
    \includegraphics[width=14cm, height=10cm]{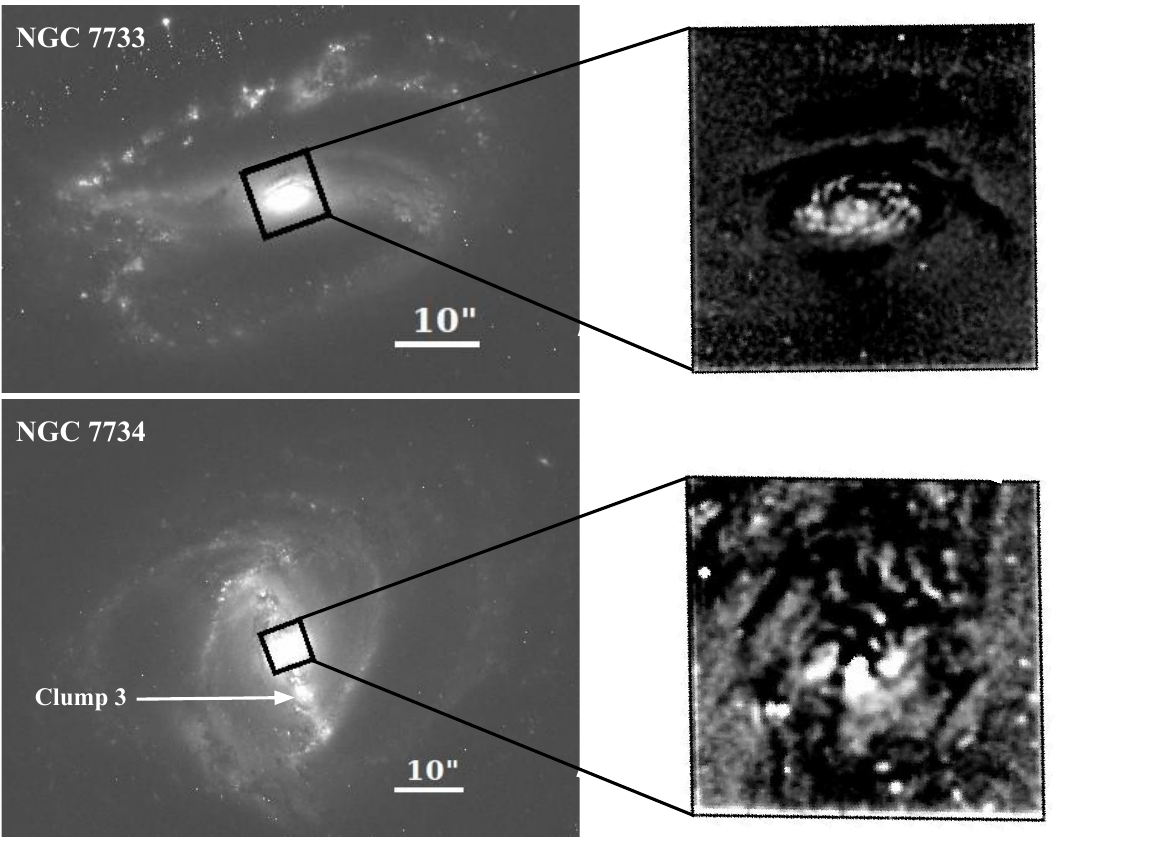}
    \caption{The $F606W$ filter images from ACS/HST of NGC 7733 (top) and NGC 7734 (bottom) are shown, along with the structure maps of the central regions for both galaxies. For NGC 7733, the structure map is zoomed in to cover the central 8$\times$8 arcsec$^{2}$, revealing a visible central disc-like structure with a semi-major axis of $1.515$ arcsec. At the distance of NGC 7733, 1 arcsec = 737 pc. In contrast, the structure map for NGC 7734 is zoomed in to cover the central 6$\times$6 arcsec$^{2}$, showing a diffuse structure in the centre. The clump present along the bar of NGC 7734 is marked here and has been discussed in Sec.~\ref{sec:Sec_4.3}. At the distance of NGC 7734, 1 arcsec = 766 pc}. North is up, and east is to the left.
    \label{fig:smap}
\end{figure*}

\begin{table*}[h]
	\centering
        \caption{Properties of three components (bulge, bar and disc) of NGC 7733 and NGC 7734 in $J$ and $K_s$ band images. $R_{e}$ is the effective radius in kpc for the bulge and bar, $R_{s}$ is the disc scale length in kpc, $M$ is absolute magnitude, $n$ is the Sersic index, $f_{bar}$ is the bar strength as measured from deprojected images and $M_{*}$ is the stellar mass, estimated for all three component using $K_s$ band mass-to-light ratio $M/L =0.6$ \citep{McGaugh14}.}
        \begin{tabular}{|l|c|cccc|cccc|ccc|r|}

        \toprule
        \multicolumn{1}{c}{} & \multicolumn{1}{c}{} & \multicolumn{4}{c}{\textbf{Bulge}} & \multicolumn{4}{c}{\textbf{Bar}} & \multicolumn{3}{c}{\textbf{Disc}} & \multicolumn{1}{c}{}\\
        \cmidrule(rl){3-6} \cmidrule(rl){7-10} \cmidrule(rl){11-13}
        
        \textbf{Object} & {Band} & $R_{e}$ & m$_{AB}$ & n &  $log(\frac{M_{*}}{M_{\odot}})$ & $R_{e}$ & m$_{AB}$ & n &  $log(\frac{M_{*}}{M_{\odot}})$ & $R_{s}$  & m$_{AB}$  &  $log(\frac{M_{*}}{M_{\odot}})$ & \textbf{$f_{bar}$}\\
         &  & (kpc) & &  &  & (kpc) & & &  &  (kpc) &  &   \\
         \hline
         \\
        NGC 7733 & $K_s$ & 1.28 & 13.66 & 2.21 & 11.65 & 11.83 & 13.91 & 0.47 & 11.55 & 29.95 & 11.45 & 12.54 & -- \\ 
          & $J$ &  1.59  & 13.72 & 2.25 & -- & 12.14  & 13.93 & 0.43 & -- & 30.01 & 10.93 & --& 0.54$\pm$0.01\\
        \\
        \hline
        \\
         NGC 7734 & $K_s$ & 1.10 & 14.79 & 1.73 & 11.47 & 7.59 & 13.73 & 0.55 & 11.87 & 8.77 & 12.12 & 12.63 & -- \\
         & $J$ & 1.15 & 14.34 & 1.39 & -- & 7.72 & 13.25 & 0.48 & -- & 8.85 & 11.51 & -- & 0.27$\pm$0.01 \\
           \\
           \hline           
        \end{tabular}
	\label{tab:Table_1}
\end{table*}


\subsection{Photometric Decomposition}\label{sec:Sec_3.2.}
NIR images are used for two-dimensional decomposition because the emission from the old stellar populations in galaxies dominates the NIR band, and dust has less of an impact on NIR emission. Therefore, NIR images are the best for modelling the light distribution in the bulge, bar, and discs of galaxies. We employed the two-dimensional (2-D) decomposition code GALFIT \citep{Peng02} to obtain measurements of structural components from NIR ($J$ and $K_s$) images of NGC 7733 and NGC 7734. GALFIT uses the Levenberg-Marquardt algorithm to minimise the $\chi^{2}$ between the observed image and the PSF-convolved model image (the sum of model components). We used two Sersic and one exponential profile to model the bulge, bar, and disc light distribution. Another Sersic profile was used to model the third galaxy, NGC 7733N, and details of the model parameters for this galaxy can be found in \citet{Yadav21}. Here, we present the model parameters for NGC 7733 and NGC 7734. The best-fit model is determined using a $\chi^{2}$ minimisation technique.
The estimated errors on the recovered parameters in GALFIT models consider only the random (Poisson) noise due to the source brightness when the model fits the data perfectly. However, in the case of galaxy fitting, the discrepancies between data and model profile include not just Poisson sources but also errors arising from PSF variation in the image, parameter degeneracy, and sky background estimation errors \citep{Peng10}. The unmodelled spiral arms cause further uncertainty on the retrieved values \citep{Barway20a}. However, \citet{Vika13, Vika14} estimate the associated errors of the fitted parameter based on the uncertainties in estimating the sky flux, which dominates the error budget. In bulge+disc decomposition, \citet{Vika14} show the uncertainties associated with bulge $n$ and $R_{e}$ can vary by up to 25\%. \citet{Kruk18} considered the similar uncertainties in their bulge+bar+disc decomposition.

Since estimating the errors caused by each of these components is non-trivial, we present the best-fit parameters for both NGC 7733 and NGC 7734 without the associated errors in Table~\ref{tab:Table_1}. The $J$-band NIR images, the model images, the residual images along with the 1-D surface brightness (hereafter - SB) profiles of NGC 7733 and NGC 7734 are shown in Fig~\ref{fig:galfit}. In the residual image of NGC 7734 (image (c) in the lower panel of Fig~\ref{fig:galfit}), we notice that some stellar contribution is present in the very central region. Also, from the 1-D SB profile (image (d) in the lower panel of Fig~\ref{fig:galfit}), we find some missing light (the model SB profile does not match the galaxy SB profile) in the very central 1 arcsec region. However, we cannot model it in our $J$-band data due to the resolution. We suspect that this missing light can be a contribution from a second bulge, which can be a classical bulge at the very centre of the galaxy. We tried to model the light distribution of NGC 7734 in HST/F606w band data using GALFIT, which did not return a good fit due to the contamination from dust and the structures in the NGC 7734's central region. We then look into the central isophotes of NGC 7734 using HST data and find that the central ($\sim$230 pc) isophotes are nearly circular compared to the outer isophotes and it is shown in Fig~\ref{fig:HST_Centre}. Studies on composite bulges \citep{Mendez14, Erwin15} have found the existence of a central classical bulge embedded in a disc-like bulge with the central isophotes becoming very round as we find in the centre of NGC 7734.

Based on the best-fitted model effective radius values ($R_e$), we found that the bulge and bar of NGC 7733 are larger than the respective components of NGC 7734. However, when taking into account the error bars as discussed earlier, both bulges could be consistent in size. Additionally, the disc of NGC 7733 is approximately twice the size of NGC 7734, as estimated from the disc scale length ($R_s$). Using the Sersic index ($n$) value, we found that NGC 7734 hosts a disc-like bulge ($1 < n < 2$) whereas NGC 7733 hosts a classical bulge ($n > 2$). Although some studies \citep{Costantin18} suggest that sersic index alone is not a valid method for identifying the nature of bulges, it is extensively used to classify the bulge type. However, NGC 7733 hosts a disc-like bulge rather than a classical bulge with an uncertainty of 25\% in $n$, which makes $n<2$. We also determined the mass of individual components in the $K_s$ band using $M/L = 0.6$ \citep{McGaugh14}. We did not perform reddening correction due to galactic extinction since it is negligible in the $K_s$ band. 

The existence of a disc-like bulge in NGC 7733 and NGC 7734 indicates that, most likely, it has been formed through secular evolution due to the presence of a bar.

\subsection{Structure Map}\label{sec:Sec_3.3}
The analysis of fine nuclear structural features in the central region of galaxies involves the use of a variety of image-processing techniques. One such technique, known as {\it Structure Map}, was introduced by \citet{Pogge02}. This method can resolve structures as small as an image's point spread function (PSF) scale. It effectively eliminates the smooth large-scale light distribution and enhances the visibility of narrow dust lanes and emission-line gas in nearby galaxies. We have applied this method to the HST/$F606W$ images of NGC 7733 and NGC 7734. The process involves convolving the image with a PSF for the PC1 camera corresponding to the filter band-pass generated using TINYTIM \citep{Krist11}. The original image is then divided by the PSF-convolved image, and the resulting ratio is further convolved with the transpose of the model PSF. Mathematically, a structure map S is defined as:
\begin{equation}
    S = [\frac{I}{I\otimes P}] P^{t}
\end{equation}
Where $I$ is the original image, $P$ is the model PSF, $P^{t}$ is the transpose of the PSF, and $\otimes$ is the convolution operator. This technique is based on the Richardson-Lucy (RL) image restoration algorithm. The structure maps for the central 8 arcsec$^{2}$ and 6 arcsec$^{2}$ square region of NGC 7733 and NGC 7734, respectively, are shown in Fig.~\ref{fig:smap}. Dark regions represent the dust-obscured areas, in contrast, bright regions correspond to either the location of heightened stellar light (for example, star-forming regions) or emission-line regions (the $F606W$ filter reveals several bright emission lines from high surface-brightness regions due to its wide bandwidth). In NGC 7733, we identified a nuclear disc-like structure that appears like a nuclear ring due to presence of dust. The semi-major axis of this nuclear disc is $\sim$ $1.116$ kpc. Conversely, the central feature in NGC 7734 resembles a disc and spiral arms-like structure in the outer parts as depicted in Fig.~\ref{fig:smap}.


\subsection{MUSE data cube analysis}\label{sec:Sec_3.4}
We utilized the wide field mode (WFM) of the MUSE IFS, which provides an archival science-ready data cube of the galaxy group NGC 7733-34. The data cube was analysed using the 3.1.0 version of Galaxy IFU Spectroscopy Tool (GIST) \citep{Bittner19}, a comprehensive scientific analysis framework for fully reduced (integral-field) spectroscopic data. GIST leverages the penalized pixel-fitting code (pPXF) \citep{Cappellari04, Cappellari17, Cappellari2022} to extract stellar line-of-sight velocity distribution (LOSVD), that is, stellar kinematics, and the Gas and Absorption Line Fitting (GandALF) \citep{Sarzi06, Falcon06} to determine the emission line fluxes and gaseous kinematics. Voronoi binning \citep{Cappellari03} was used for spatially binning the data within the wavelength range $480 - 580$ nm to achieve a S/N of $30$. Spaxels with S/N greater than $30$ remain unbinned, while those with S/N less than three were excluded to eliminate noisy data before binning. Additionally, all spectra were corrected for galactic extinction values in the direction of the target. It is important to keep in mind that the selected wavelength range includes a potentially strong emission line ($H\beta$). However, in the NGC 7733-34 group, $H\beta$ emission is significantly low, thus the Voronoi binning will not be substantially affected by the selected wavelength range. 

For the analysis of stellar kinematics, pPXF utilizes a non-negative linear combination of template spectral libraries convolved with the LOSVD in pixel space. The method employs a least square minimisation technique to determine the best fit LOSVD parameters. In the current analysis, to prevent the contamination from the redder wavelength, the fitting of the spectra is carried out between the wavelength range $480-580$ nm, as suggested by \citet{Bittner19, Bittner21}. During the fit, prominent emission lines within this range are masked. An 8th-order multiplicative Legendre polynomial is incorporated into the fit to address any minor differences in the continuum between observed and template spectra. Taking these best-fit values into account, spatial maps of the stellar velocity ($V$) in the galaxy's rest frame, velocity dispersion ($\sigma$), and higher order Gauss-Hermite moments ($h_{3}$ and $h_{4}$) of the galaxies are constructed. In this work, we are using V, $\sigma$ and $h_{3}$ maps for further discussions. The $h_{4}$ moment is not used because it does not provide relevant information for this work; thus, we haven't included a corresponding map of $h_{4}$ moment.

\subsection{Detection of SFCs and SFR estimation}\label{sec:Sec_3.5}
In the science-ready $FUV$ and narrow band $H\alpha$ images of NGC 7733 and NGC 7734, the star-forming regions appear as bright clumps. To identify these clumps, we utilized the Source Extractor Software (SExtractor) \citep{Bertin96}. This software takes a threshold value as input to detect the source, with pixels exceeding the threshold count identified as a source. For the $FUV$ image, we set the detection threshold at 1.5$\sigma$, and for the $H\alpha$ image, it is $3 \sigma$, where $\sigma$ represents the global background noise. In the image of NGC 7733, we detected 71 regions in $FUV$ and 53 regions in $H\alpha$. Similarly, in the image of NGC 7734, we ascertained 181 regions in $FUV$ and 64 regions in $H\alpha$. After correcting for instrumental resolution, we found 41 $FUV$ and 51 $H\alpha$ regions in the image of NGC 7733, with three of the regions associated with nuclear emission. For the image of NGC 7734, we obtained 73 $FUV$ and 51 $H\alpha$ regions. Among these, 11 $FUV$ and 4 $H\alpha$ regions are located along the bar. 

In order to conduct aperture photometry on the identified sources, we employed Photutils, which is a Python astropy package\footnote{https://photutils.readthedocs.io/en/stable/aperture.html}. The star-forming regions represent a group of young, vibrant stars, and the magnitude calculated from these regions is influenced by both galactic and internal extinction. We determined galactic extinction in $UV$ and $H\alpha$ by utilizing the extinction law outlined by \citet{Cardelli89} -
\begin{equation}
    <\frac{A(\lambda)}{A(V)}> = a(x)+\frac{b(x)}{R_{v}}
    \label{eq:3}
\end{equation}
Here, x is the wave number, a(x) and b(x) were evaluated from \citet{Cardelli89}, $R_{v} = 3.1$ for Milky Way  and A(V) = 0.073 for NGC 7733 and 0.074 for NGC 7734.

For star-forming disc galaxies, \citet{Kennicutt98} suggested  $H\alpha$ extinction, $A_{H\alpha} = 1.1$ mag. To correct for internal extinction we used $A_{H\alpha} = 1.1$ mag and $A_{FUV}/A_{H\alpha} = 3.6$ \citep{Leroy08}. We derived the SFR in the $FUV$ filter using the following relation \citep{Leroy08} -
\begin{equation}
    SFR_{FUV}(M_\odot\ yr^{-1}) = 0.68\times10^{-28}L_{\nu}(FUV).
	\label{eq:4}
\end{equation}
To measure the SFR of the $H\alpha$ star-forming regions we used the following relation \citep{Calzetti07} -
\begin{equation}
    SFR_{H\alpha}(M_\odot\ yr^{-1}) = 5.3\times10^{-42}L_{H\alpha}(erg/s).
	\label{eq:5}
\end{equation}
The luminosity in both the equations (Eqs.~\ref{eq:4} and ~\ref{eq:5}) is the intrinsic luminosity.

\section{Results and  Discussion}\label{sec:Sec_4}
\subsection{Stellar Kinematics of the bulge}\label{sec:Sec_4.1}
In this section, we examine the signatures present in the kinematic maps obtained from MUSE data, which indicate the existence of various structural components. The higher-order $h_{3}$ moment which quantifies the skewness of LOSVD is crucial to characterize the stellar orbital structure. 
\begin{figure*}
	\includegraphics[width=\textwidth]{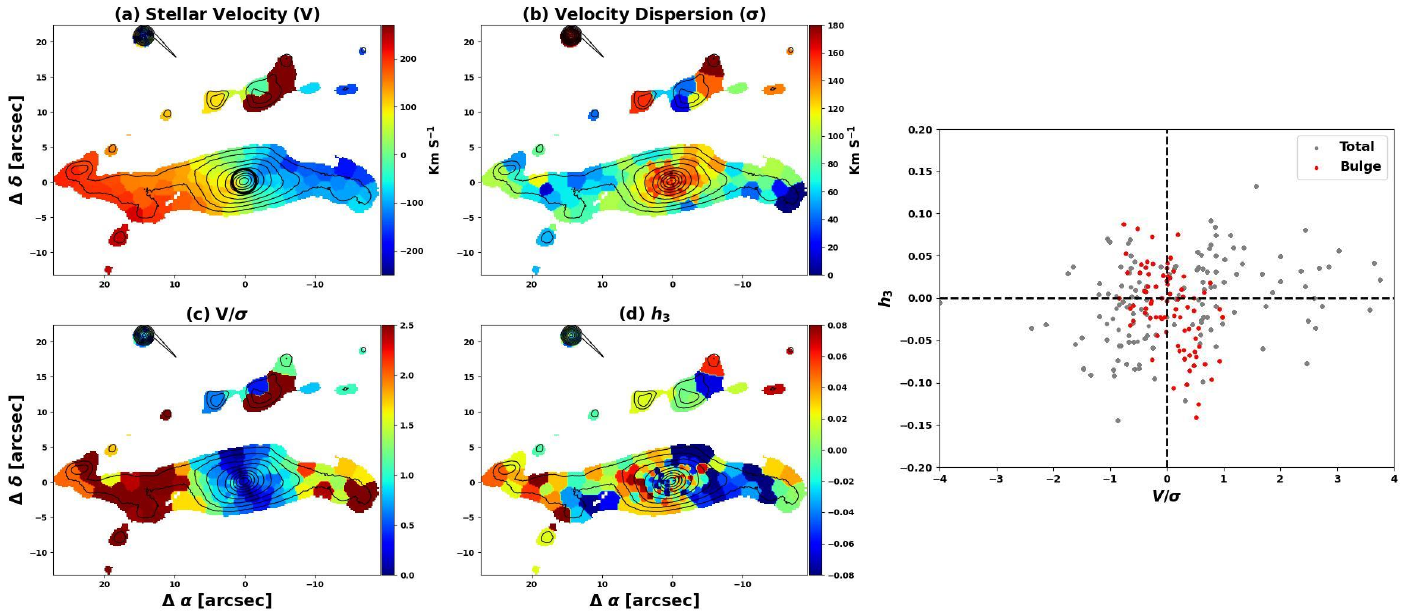}
        \caption{\textbf{Left Panel}: Stellar Kinematic maps of NGC 7733, showing stellar velocity ($V$), velocity dispersion ($\sigma$), $V/\sigma$, and $h_{3}$ as indicated. The isophotes, equally spaced in 0.15 mag steps, are derived from MUSE intensity maps. Colour bars beside each map indicate the parameter ranges. The black circle on the velocity map marks the bulge, with a radius equal to the bulge effective radius from photometric decomposition using the $K_s$ band NIR image, as detailed in Table~\ref{tab:Table_1}. 1 arcsec = 737 pc. North is up, and east is to the left. \textbf{Right Panel}: $h_{3}$ vs $V/\sigma$ plot for each Voronoi bin. Red circles denote bins within the bulge effective radius (inside the black circle on the velocity map), while grey points represent bins for the entire galaxy.}        
        \label{fig:stkin_33}
\end{figure*}
\begin{figure*}
	\includegraphics[width=\textwidth]{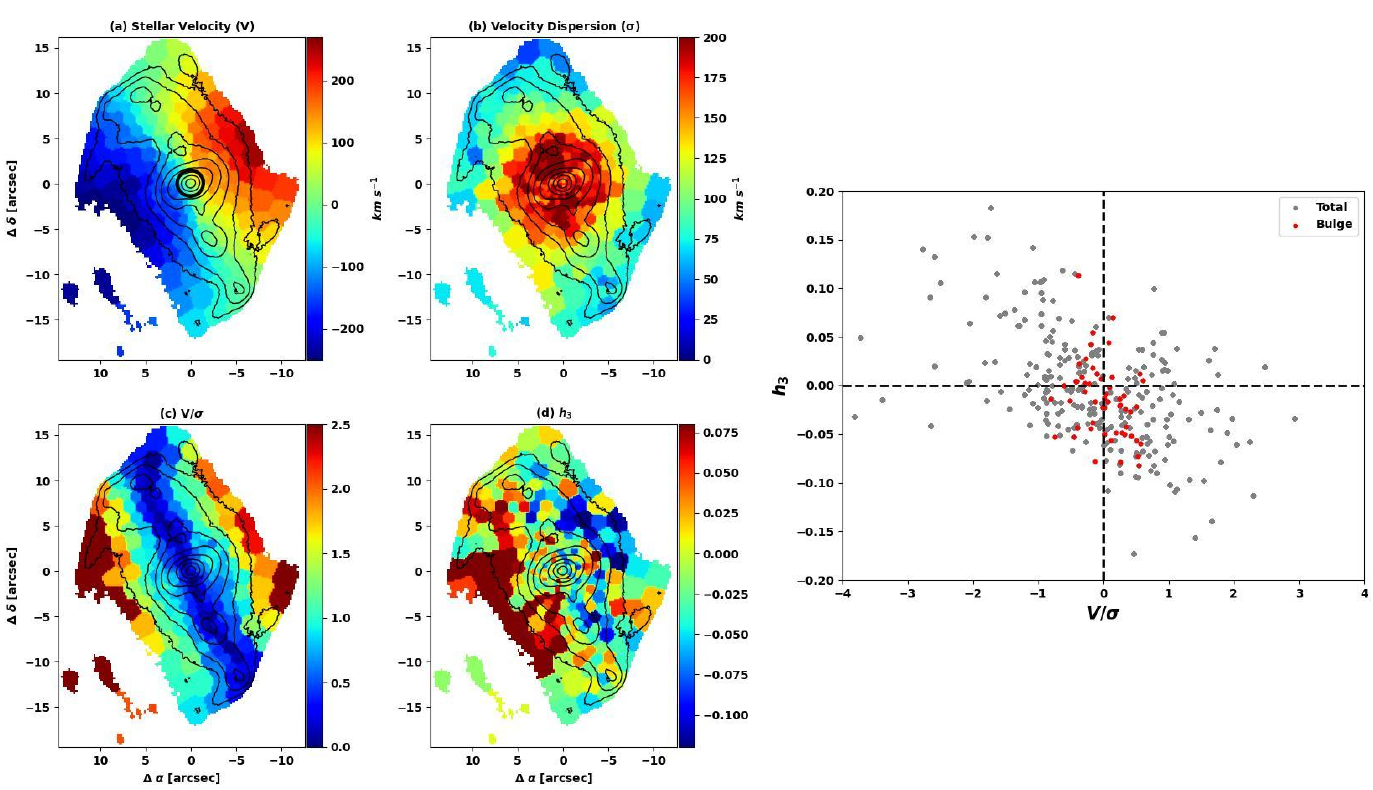}
        \caption{\textbf{Left panel}: Stellar Kinematic maps of NGC 7734, displaying stellar velocity ($V$), velocity dispersion ($\sigma$), $V/\sigma$, and $h_{3}$, as labelled. The isophotes are evenly spaced in intervals of 0.15 mag, derived from MUSE intensity maps. Colour bars next to each map indicate the range of values for each parameter. The black circle on the velocity map denotes the bulge, with a radius equal to the bulge effective radius obtained from photometric decomposition using $K_s$ band NIR imaging, detailed in Table~\ref{tab:Table_1}. 1 arcsec = 766 pc. North is up, and east is to the left. \textbf{Right panel}: $h_{3}$ vs $V/\sigma$ plot for each Voronoi bin. The red circles represent bins within the bulge effective radius (inside the black circle on the velocity map), while grey points denote bins across the entire galaxy. }        
        \label{fig:stkin_34}
\end{figure*}
\subsubsection{NGC 7733}\label{sec:Sec_4.1.1}
In the left panel of Fig.~\ref{fig:stkin_33}, the maps of $V$, $\sigma$, $V/\sigma$, and $h_{3}$ moment of NGC 7733 are displayed. The contours, derived from the intensity maps from MUSE, are equally spaced in steps of 0.2 mag. The stellar velocity map exhibits a regular rotation pattern, with the north-east region of the galaxy showing a red-shifted velocity and the north-west region displaying a blue-shifted velocity. The black circle represents the bulge, with a radius equal to the bulge's effective radius estimated from GALFIT modeling on the $K_s$ band image (see, Table~\ref{tab:Table_1}). A distinct kinematic component with enhanced velocity is observable in the central region of 2 arcsec radii. Additionally, the northern spiral arm region shows a velocity different from that of NGC 7733, which has already been studied by \citet{Yadav21}. This discrepancy is attributed to the presence of the other galaxy, NGC 7733N, which is moving with a velocity of $\sim$ 650 km s$^{-1}$ with respect to NGC 7733 \citep{Yadav21}. 

The velocity dispersion map shows an increase in $\sigma$ within the galaxy's central region, spanning approximately 10 arcsec. The median of the velocity dispersion within the effective radius of the bulge is 148 $\pm$ 14 km s$^{-1}$. Additionally, a high-velocity dispersion is observed in the knot located in the northern spiral arm, corresponding to the neighbouring galaxy NGC 7733N. By utilizing the $V$ and $\sigma$ maps, we generated a map displaying the ratio of line-of-sight velocity to velocity dispersion ($V/\sigma$). The $V/\sigma$ values mainly fall within the range of approximately -3 to 3, but within the bulge effective radius, they are constrained to around -1 to 1. Furthermore, we identified an anti-correlation between the $V$ and $h_{3}$ moment within the bulge's effective radius, with $h_{3}$ values ranging mostly from -0.12 to 0.12. This anti-correlation suggests the presence of near-circular stellar orbits, whereas a correlation between them indicates eccentric stellar orbits. This finding of anti-correlation between $V$ and $h_{3}$ may be linked to the nuclear disc identified using the structure map on the Hubble Space Telescope image and is depicted in Fig.~\ref{fig:stkin_33} \citep{Gadotti15, Bittner19}. 

In order to provide additional kinematic evidence supporting the presence of a nuclear disc, we used a $V/\sigma$ vs $h_{3}$ map to graph their values for all the bins, which is displayed in the right panel of Fig.~\ref{fig:stkin_33}. The grey dots represent the values for each bin of the galaxy, while the red dots indicate the values corresponding to the bins inside the bulge effective radius. The $V/\sigma$-$h_{3}$ values do not reveal a strong correlation or anti-correlation for the entire galaxy. However, they show a significant anti-correlation for the bins within the bulge effective radius. The Spearman correlation coefficient for the bulge (red dots) is -0.55, indicating the anti-correlation and providing further evidence for the presence of circular orbits in the central region, thereby supporting the existence of nuclear structure. Previous studies have reported the formation of nuclear structures in the vicinity of inner Lindblad resonances \citep{Knapen95, Comeron10} through bar-stimulated gas inflow \citep{Combes85}. The existence of nuclear discs can also be supported by stellar kinematics. An anti-correlation between $V$ and $h_{3}$ aligns with circular motion \citep{Gadotti15}, further confirming the presence of nuclear disc structure. We found that the semi-major axis of the disc aligns with the major axis of bar. 

\subsubsection{NGC 7734}\label{sec:Sec_4.1.2}
The left panel of Fig.~\ref{fig:stkin_34} shows the $V$, $\sigma$, $V/\sigma$, and $h_{3}$ moment maps of NGC 7734. The contours are equally spaced in steps of 0.15 mag and derived from the MUSE intensity maps. The stellar velocity map displays a regular rotation pattern, with the north-western region of the galaxy having a red-shifted velocity and the south-eastern region having a blue-shifted velocity. The black circle denotes the bulge, and its radius is equal to the radius of the bulge effective radius estimated from GALFIT modeling on $K_s$ band image. It is given in Table~\ref{tab:Table_1}. The $\sigma$ map reveals the enhancement of $\sigma$ in the galaxy's central region with a radius of 5 arcsec. The median of velocity dispersion within the bulge effective radius is 179 $\pm$ 16 km/s. The higher $\sigma$ value means that the stars are generally dynamically hotter, as found in star-forming galaxies \citep{oh22}. 

We produced the ratio of LOS velocity and velocity dispersion ($V/\sigma$) map using the $V$ and $\sigma$ maps. The $V/\sigma$ value mostly ranges between $\sim$ -3 to 3, whereas it lies between $\sim$ -1 to 1 for the bulge effective radius. We found an anti-correlation between V and $h_{3}$ moment within the effective radius of the bulge. The $h_{3}$ value mostly has a range between $-0.12$ and $0.12$. Similar to NGC 7733, we used $V/\sigma$ vs $h_{3}$ map to plot their values for all the bins and has been shown in the right panel of Fig.~\ref{fig:stkin_34}. Here, the grey dots present the values for each bin of the galaxy, whereas the red dots represent the value corresponding to the bins inside the bulge effective radius. We found the Spearman correlation coefficient for the bulge (red dots) is -0.38, which suggests a loose anti-correlation and supports the presence of a circular orbit in the central region. However, we notice a disc and spiral arms-like structure in the central region from the structure map on HST image, resulting in slight $V/\sigma$ - $h_3$ anti-correlation. Although, from the photometric decomposition, we found that NGC 7734 hosts a disc-like bulge, a secular built structure that a stable circular orbit can represent, we also see the presence of stellar light contribution in the very centre of the galaxy along with the central circular isophotes as discussed in Sec~\ref{sec:Sec_3.2.}. Taking all these factors into account, it can be concluded that NGC 7734 exhibits multiple central structures, including a second bulge. Specifically, the galaxy contains a classical bulge with a radius of approximately $\sim$230 pc, alongside a disc-like bulge with an effective radius of $\sim$1.15 kpc in the $J$-band. Furthermore, the high-velocity dispersion observed in the central region provides additional evidence supporting the existence of a classical bulge at the centre of NGC 7734 \citep{Erwin15}. 
\begin{figure*}
        \hspace{0.5cm}
	\includegraphics[width=18cm,height=10cm]{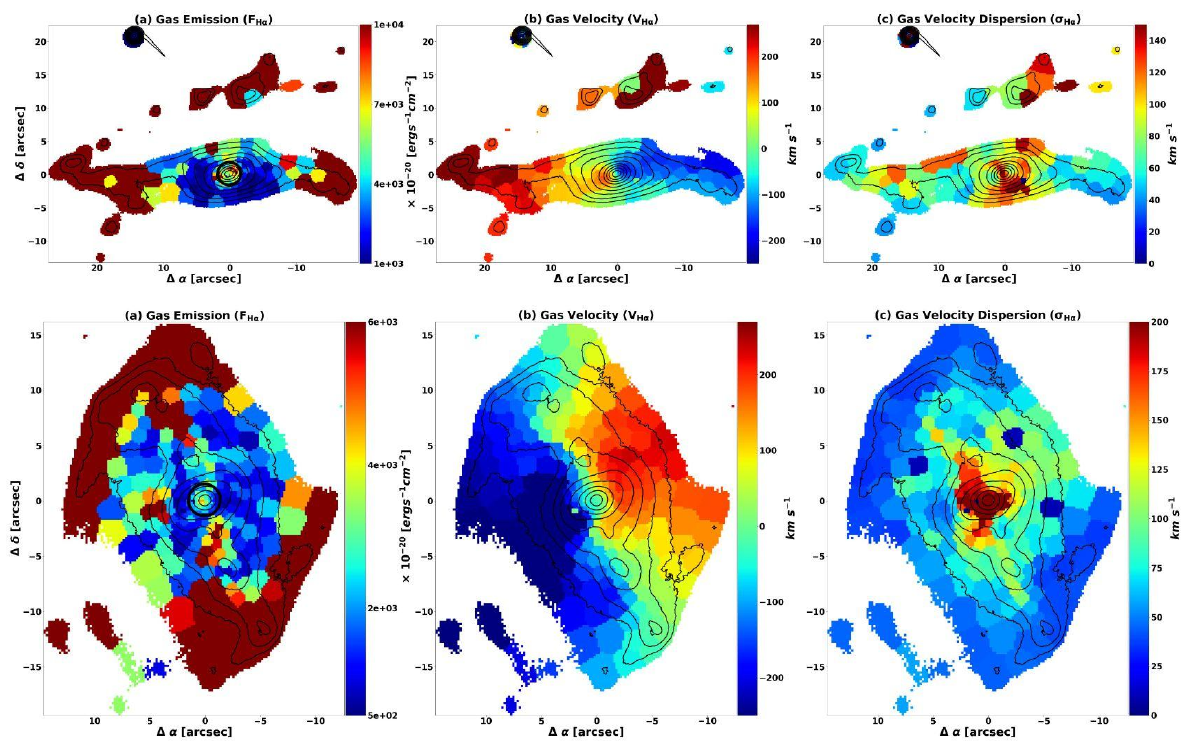}
    \caption{Gas ($H\alpha$) intensity \& kinematic maps of NGC 7733 (top panel) and NGC 7734 (bottom panel). (a) represents the gas intensity map, (b) shows the gas velocity map, and (c) displays the gas velocity dispersion map for NGC 7733 (top panel) and NGC 7734 (bottom panel). The isophotes are equally spaced in intervals of 0.15 mag and are derived from the MUSE intensity maps. The colour bars beside each map indicate the range of the parameters. The black circle on the intensity map marks the bulge, with its radius equal to the bulge effective radius derived from photometric decomposition using $K_s$ band NIR image, as specified in Table~\ref{tab:Table_1}. At the distance of NGC 7733, 1 arcsec = 737 pc. Whereas, at the distance of NGC 7734, 1 arcsec = 766 pc. North is up, and east is to the left.}
    \label{fig:gaskin}
\end{figure*} 

\subsection{Gas Kinematics of bulge}\label{sec:Sec_4.2}
To measure emission line properties such as emission line fluxes, gas kinematics, and emission-free continuum spectra, GIST utilizes the GandALF module \citep{Sarzi06, Falcon06}. Each user-specified emission line is treated as a Gaussian template additional to the stellar templates. The best-fitting gaseous properties are estimated using a linear combination of these emission line templates with the spectral library template. We have generated gas ($H\alpha$) velocity and velocity dispersion maps for NGC 7733 (top) and NGC 7734 (bottom), as shown in Fig.~\ref{fig:gaskin}. 

\subsubsection{NGC 7733}\label{sec:Sec_4.2.1}
The gas velocity map (image (a): top left) for NGC 7733 exhibits a similar pattern to the stellar velocity. Specifically, the north-eastern region of the galaxy displays a red-shifted velocity, while the north-western region shows a blue-shifted velocity. Additionally, the central region demonstrates a slightly enhanced gas velocity component compared to the stellar velocity. Moving on to the gas velocity dispersion map (image (b): top right), it reveals a higher velocity dispersion in the central few arcsecs of the galaxy. Within the bulge's effective radius, the gas velocity dispersion measures 114 $\pm$ 10 km s$^{-1}$. The gas velocity dispersion in the centre is lower compared to the stellar velocity dispersion. The presence of the nuclear disc, a rotationally supported system, contributes to this lower dispersion.

Although NGC 7733 hosts an AGN of Seyfert 2 at its centre, it also exhibits an Extended Narrow Line Region (ENLR) towards the north-eastern area \citep{Yadav21}. Studies indicate that gas velocity dispersion is strongly influenced by the power source of AGN outflows, which heat the gas while maintaining dynamical equilibrium \citep{oh22}. However, the case of NGC 7733 presents a conflicting scenario, possibly due to the presence of a rotationally supported system, the nuclear disc, where stellar and gas velocity dispersion lowers beyond the central few arcsec regions. 

This change in stellar and gas velocity dispersion could be an outcome of central SF as discussed by \citet{Wozniak03}. In NGC 7733, we can see the presence of central $H\alpha$ and $FUV$ emissions (see, the $FUV$ and $H\alpha$ image in the upper panel of Fig. 1. of \citet{Yadav21}), which is associated to the nuclear disc. 

\subsubsection{NGC 7734}\label{sec:Sec_4.2.2}
The gas velocity map (image (a): bottom left) of NGC 7734 exhibits a similar pattern to the stellar velocity. In the north-western region of the galaxy, there is a red-shifted velocity, while the south-eastern region shows a blue-shifted velocity. The gas velocity in the centre is slightly higher than the stellar velocity. Furthermore, the gas velocity dispersion (image (b): bottom right) in the central few arcsec of the galaxy is notably high. The dispersion measures 224 $\pm$ 22 km s$^{-1}$ within the bulge's effective radius. Notably, the gas velocity at the centre surpasses the stellar velocity dispersion. The presence of high gas velocity dispersion is interesting to notice. NGC 7734 hosts LINER nuclei, and the majority of local AGN exhibit spectra dominated by low-ionisation emissions such as [OI], [NII], and [SII] and are classified as LINERs \citep{Heckman80, Cid10}. The prevailing view is that these LINERs are predominantly powered by hot, evolved stars \citep{Belfiore15}. It suggests both evolved stars and AGN activity may collaboratively contribute to the enhanced gas velocity dispersion observed in the centre of NGC 7734. It implies that the increased gas velocity dispersion at the galaxy's core can be attributed to AGN activity. \citet{oh22} found that AGN and LINER regions, especially those concentrated in the central parts of galaxies, tend to produce dynamically hotter gas kinematics, characterised by large velocity dispersion, compared to regions dominated by active SF. Furthermore, studies have identified a positive correlation between gas velocity dispersion and the diagnostic emission line ratios on the BPT diagram \citep{Ibero06}. This correlation is indicative of shock-driven ionisation processes. Additionally, the existence of tidally induced large-scale gas flows, often a result of galaxy mergers or interactions, can be the source of these shocks \citep{Ibero06, Ibero10}. For NGC 7733-34, which is a closely interacting galaxy system, the observation of several small galaxies within the field suggests that this group may have experienced numerous minor mergers or fly-by interactions. The presence of NGC 7733N in the northern arm of NGC 7733 further corroborates this hypothesis \citep{Yadav21}. 


\subsection{Star formation along the bar of NGC 7734}\label{sec:Sec_4.3}
In our investigation, we have conducted a multi-wavelength analysis to examine the occurrence of $H\alpha$ and $FUV$ emission along the bar of the barred spiral galaxy NGC 7734 in the triple AGN group NGC 7733-34. A recent study by \citet{Yadav23} examined the SF in NGC 7733 and NGC 7734 using $FUV$ data from UVIT and $H\alpha$ data from MUSE. Our measured mean $\Sigma_{SFR}$ aligns with their estimated values for both galaxies. Additionally, they documented the presence of $H\alpha$ emission along the bar of NGC 7734. 
\begin{figure*}
	\includegraphics[width=\textwidth]{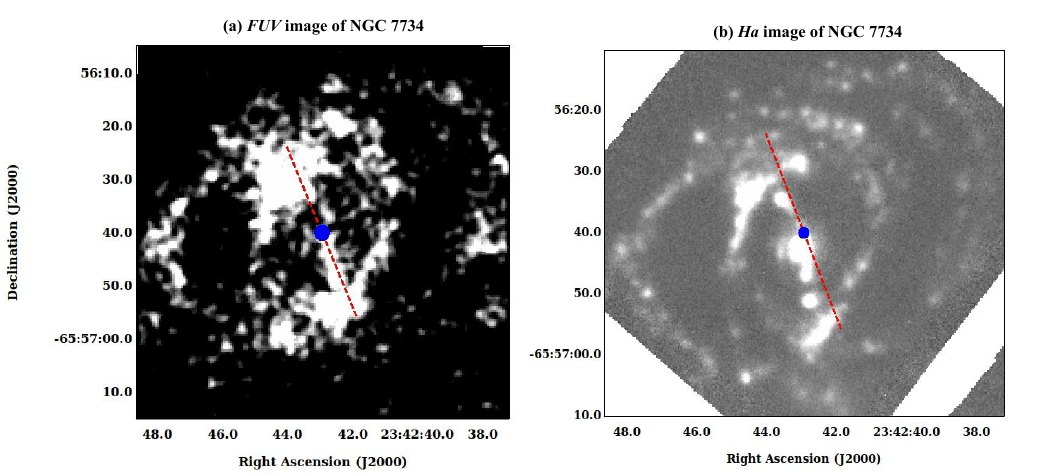}
    \caption{\textbf{Left}:$FUV$ image of NGC 7734 from UVIT, \textbf{Right}: $H\alpha$ narrow band image of NGC 7734 from MUSE. The red dashed line in each image is the bar PA as estimated from isophotal analysis on $J$-band image and the blue solid circle of radius 1 arcsec represents the photometric centre of the galaxy. Each image frame covers an area of $70 \times 70$arcsec$^2$. North is up, and east is to the left.}
    \label{fig:pa}
\end{figure*} 
We conducted a similar analysis on both the $FUV$ and $H\alpha$ images of NGC 7734. Remarkably, we observed the presence of $FUV$ emission along with $H\alpha$ in the bar region of NGC 7734. We identified and characterised all the SFCs in the $FUV$ and $H\alpha$ images and calculated their $\Sigma_{SFR}$. We categorized the SFCS into three regions: 1) Central region, which encompasses the central part of the galaxy, including the bar and bar ends, 2) Inner ring and 3) Outer ring. The detailed discussion on the SF in NGC 7734 is given is Appendix~\ref{app:Appendix_A}. The number of detected clumps along with the median value of log $\Sigma_{SFR}$ for the three regions as described above are presented in Table~\ref{tab:Table_A1}. 

In previous research, the impact of the bar on central SF has been documented \citep{Combes85}. Understanding SF along the bar is crucial for gaining insights into the role of the bar in galaxy evolution. In a recent study, \citet{Fraser-McKelvie20} identified the presence of $H\alpha$ emissions along the bar and at the bar ends in a sample of MaNGA galaxies, particularly those in low-density environments. Their findings indicate that the $H\alpha$ bar leads the stellar bar, as evidenced by measuring the respective $H\alpha$ and stellar bar position angles. 

In Fig.~\ref{fig:pa}, we present the UVIT $FUV$ image (left panel) and MUSE $H\alpha$ image (right panel) of NGC 7734. We have overlaid the $J$ band bar PA on both the $FUV$ and $H\alpha$ images passing through the galaxies' photometric centre, which we determined from the isophotal analysis. The PA of the bar is measured from isophotal analysis on $J$ band image as discussed in Sec.~\ref{sec:Sec_3.1}. The estimated bar PA of NGC 7734 is 22.48\textdegree $\pm$ 0.79\textdegree. We found that there is an observable difference in the position angle between the stellar bar and $H\alpha$ along the stellar bar, whereas the $FUV$ bar is primarily aligned with the stellar bar. The presence of $H\alpha$ emission along the bar is intriguing as it suggests recent SF (around 10 million years ago). Additionally, the coexistence of $FUV$ and $H\alpha$ emissions along the bar indicates the presence of young stars in this region (approximately 100 million years old). It is well known that bar-induced gas inflow triggers SF in the central regions of galaxies, and it has been observed that barred galaxies tend to have higher gas concentrations than their unbarred counterparts, which correlates with increased rates of central SF (see, \citet{Sakamoto99, Jogee05}). 

\begin{figure*}
	\includegraphics[width=\textwidth]{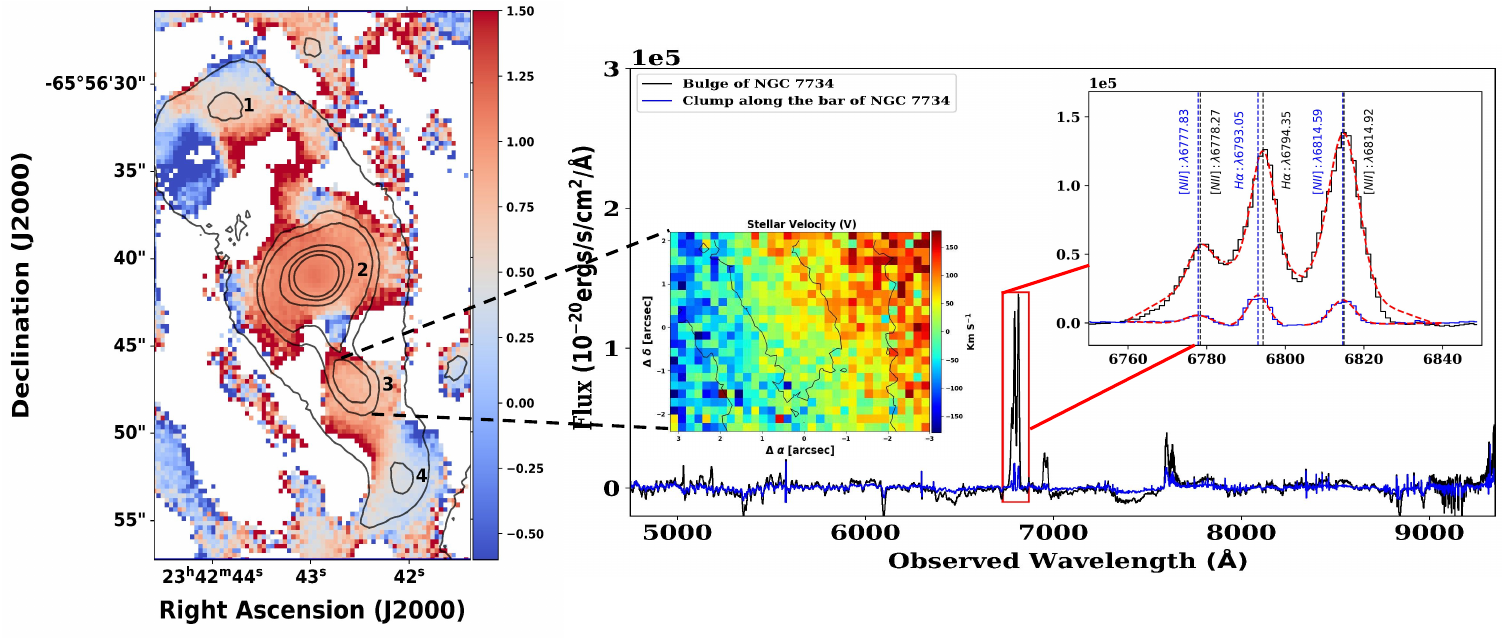}
    \caption{\textbf{Left panel:} The $g-r$ colour image of NGC 7734 bar with $g$-band contours overlaid. The clumpy nature of the bar in the optical image is marked here. Clump 2 represents the centre of NGC 7734, whereas the red clump present along the bar is highlighted as clump 3. The other two clumps (clump 1 $\And$ clump 4) are present at the bar ends. The blank spaces in the $g-r$ colour image are due to less S/N in those regions. \textbf{Right panel:} The spectra corresponding to the bulge (black) and clump 3 (blue) as shown in left panel are presented here. The spectra around [NII] and $H\alpha$ line are enhanced and are shown at the right corner. All the lines are modelled using Gaussian profiles and the observed wavelength of each line is mentioned. These lines are used to estimate the redshift of the clump present along the bar. The LOS stellar velocity map of clump 3 is also shown in the right panel. We derived the LOS velocity using the GIST Pipeline as discussed in Sec.~\ref{sec:Sec_3.4}.}
    \label{fig:color}
\end{figure*}

Additionally, the $g-r$ colour image of NGC 7734 bar is displayed in the left panel of Fig.~\ref{fig:color}, with contours taken from the $g$-band image. A detailed examination of the bar reveals clumpy stellar light within the optical contour, which we have marked as (1, 2, 3, and 4, respectively). Clumps 1 and 4 are located at the northern and southern ends of the bar respectively, while clump 2 represents the central region, identified as a disc-like bulge based on the Sersic index. Clump 3 is the clump present along the bar on the southern side of the bulge, it has also been marked in Fig.~\ref{fig:HST} and bottom panel of Fig.~\ref{fig:smap}. The presence of a clumpy structure, particularly clump 3, along the bar in the optical image, is noteworthy. We observe $FUV$ emission around the clumpy region in the left panel of Fig.~\ref{fig:pa}, while the $H\alpha$ gas is not present there, having instead moved eastward.

We suspect that clump 3 can be a compact galaxy projected in a way that it appears to be part of the bar. In order to confirm this, we extracted the spectra from the MUSE cube by taking a circular aperture with a radius of 2.26 arcsec around the clump as well as around the centre of NGC 7734. The continuum subtracted observed spectra for bulge and clump 3 are shown by black and blue lines respectively in the right panel of Fig~\ref{fig:color}. We zoomed in the spectrum around the H$\alpha$ and [NII] doublet emission lines. We fitted Gaussian profiles to all the three spectral lines and the fitted spectra are shown in red. We also plotted the peaks of the emission lines along with their respective observed wavelength. We estimated the red-shift using $H\alpha$ and [NII] lines from the MUSE data cube and found it to be 0.03508 $\pm$ 0.00003, which is similar to the red-shift of the host galaxy (z = 0.03527 $\pm$ 0.00006) \citep{Yadav21}. Thus, the clump is moving with a velocity of 57 $km$ $s^{-1}$ with respect to the galaxy NGC 7734. So, the redshift measurement does not support our assumption. We are unable to detect any velocity gradient surrounding the clump in our stellar velocity map (see, Fig.~\ref{fig:stkin_34}), similar to \citet{Yadav21}'s discovery of NGC 7733N.

However, the size and prominence of clump 3 in the optical images enforces us to estimate its mass. We measured its mass using the $g-r$ colour following the relation of \citet{Bell03}. By visually inspecting a circular region with a radius of 2.26 arcsec (approximately 1.73 kpc at the distance of NGC 7734), we determined the galactic extinction corrected colour and, hence, the mass to be (3.37 $\pm$ 0.44) $\times$ 10$^{7}$ $M_{\odot}$. Additionally, it appears red in the $g-r$ colour map with a colour value of 0.826 $\pm$ 0.05, indicating that it is not merely a star-forming clump.

Studies of compact elliptical galaxies (cEs) suggest that they have stellar masses of $10^{8} M_{\odot} - 10^{10} M_{\odot}$ with sizes in the range of 100 - 900 pc and thus they have very high stellar density \citep{Faber73, Choi02}. These low mass cEs are preferentially observed near the giant galaxies (like the presence of M32-like compact object in the close proximity of the M31 galaxy) \citep{Nieto87}. The kinematical study of cEs by \citet{Ferre21} shows that the LOS stellar velocity of these galaxies have a maximum value of around 20-60 $km$ $s^{-1}$. Utilizing the GIST workflow on the MUSE data cube as discussed in Sec.~\ref{sec:Sec_3.4}, we derived the LOS stellar kinematics of clump 3. The stellar velocity map is shown in the right panel of Fig.~\ref{fig:color}. We estimated the median value of LOS velocity of clump 3 by considering a circular region of radius 2.26 arcsec. The velocity value is 26 $\pm$ 4 $km$$s^{-1}$. 

Although the estimated stellar mass of clump 3 is slightly less compared to the CEs, it falls in the regime of ultra-compact dwarf galaxies (UCDs). The UCDs have stellar masses of $10^{6} M_{\odot} - 10^{8} M_{\odot}$ with a projected radius of up to 20 pc, making them the most compact galaxies in the universe \citep{Drinkwater00, Brodie11}. In our analysis, the stellar mass of clump 3 falls in the regime of UCDs, but its size is large, rejecting its possibility of being a UCD. Thus, based on the mass and size of the clump, we are speculating that it can be a compact galaxy. Also, the LOS velocity measurement can be contaminated due to the presence of the bar in NGC7734, which is along the line of nodes of the galaxy, so the LOS velocity measurement associated with clump 3 is limited. It will represent the upper limit of the velocity of clump 3.
Hence, the mass and the LOS velocity measurement of clump 3 follow the classification of cEs. So there is a possibility that clump 3 is a compact elliptical galaxy (cE) and hence it could be the fourth galaxy in this interacting group. We named it as NGC 7734S. This interaction seems to be the primary factor driving the continuous formation of stars in the central region. Additionally, the effect of this interaction may affect the kinematics (both stellar and gas) of the galaxy centre (despite the galaxy containing a disc-like bulge) and this has been discussed in Sec.~\ref{sec:Sec_4.1.2} and Sec.~\ref{sec:Sec_4.2.2}. 
\subsection{Bar quenching in AGN host galaxy}\label{sec:Sec_4.4}
The interaction between galaxies can result in gas exchange and gas inflow towards the central region, triggering the formation of central stars \citep{DiMatteo07, Hopkins08}. While this interaction can also trigger an AGN, it is not the sole cause of AGN activity \citep{Ellison11}. Some studies have found evidence of mergers associated with AGN activity \citep{Surace98, Smirnova06, Combes09}, while others have claimed that statistically there is no evidence of a merger associated with AGN activity, with a similar percentage found in both active and inactive galaxies \citep{Coldwell06}. Conversely, AGN feedback can lead to the cessation of SF in the central region. \citet{Jahan-Miri01} have shown that the nuclei of both galaxies in the NGC 7733-34 galaxy group are redder compared to the other parts of the galaxies, indicating the absence of nuclear starburst activity in the pair. 

In this study, we investigate the SF along the bar in NGC 7734 and bar quenching and/or inside-out quenching resulting from AGN activity in NGC 7733 of this interacting system of galaxies. The feedback from an active AGN depletes gas from the central region, leading to the cessation of SF. Studies by \citet{Ellison18, Bluck20, Lammers22} examined central SF suppression (inside-out quenching) in a large sample of MaNGA galaxies. They concluded that AGN feedback suppresses central (kpc scale) SF in nearby galaxies, it can also cause SF quenching along the bar, known as bar quenching. Additionally, \citet{Lammers22} demonstrated that AGN feedback could not drive galaxy-wide quenching in our nearby universe, instead leaving galaxies in the green valley. Feedback from the active AGN heats the interstellar medium (ISM), preventing the gas from cooling and ceasing the gas inflow towards the centre, suppressing the SF along the bar. This result provides direct observational evidence of AGN feedback suppressing SF. Our findings indicate the absence of SF in the bar of NGC 7733, which appears redder. Notably, NGC 7733 hosts a Seyfert 2 AGN as identified by \citet{Yadav21}. Ongoing interaction with NGC 7733N and NGC 7734 has led to continued SF in the outer ring/spiral arm of NGC 7733. The SFR is higher towards the northern spiral arm, likely due to the presence of NGC 7733N. The lack of SF along the bar could be attributed to the negative feedback from the active AGN. 

Strong bars are evident and long and are more frequently associated with the presence of AGNs than weak bars, which are typically smaller and fainter \citep{Garland24}. Observations suggest that strong bars are more commonly found in quiescent galaxies, contrasting with the weaker bars, indicating that either the presence of a strong bar contributes to the cessation of star formation or that strong bars more readily form in quiescent environments \citep{Geron21}. A comparison of bar properties in NGC 7733 and NGC 7734 reveals that the bar in NGC 7733 is both longer and stronger. A comprehensive analysis of a sample of interacting galaxies hosting bars and AGNs is necessary to explore the relationship between bar length and strength, star formation, and AGN activity in barred galaxies undergoing close interaction. Such study then can be compared with isolated barred galaxies \citet{Guo19}. It has been reported that the gas funnelling by the bar towards the centre increases the central SF as well as increases the gas consumption from the disc, and the galaxy turns into quenched from star-forming \citep{Kormendy04, Masters10}. \citet{Geron21} have shown that stronger bars help to quench the host galaxies through secular evolution. Our system is interacting, and we found that NGC 7733 hosts a strong and large bar with SF in the spiral arm and inside-out quenching/bar quenching. Despite having a strong bar, the $H\alpha$ and $FUV$ emission along the bar is absent in NGC 7733. Thus, a negative feedback of AGN can be a possible scenario of bar quenching in NGC 7733.


\section{Summary}\label{sec:Sec_5}
In our study, we utilized the MUSE, HST/ACS and DECaLS archival data in conjunction with observations from the UVIT and IRSF to analyse the stellar kinematics, SF and ionised gas characteristics in the central region as well as along the stellar bar of the triple AGN and interacting system NGC 7733-34. The key findings of our investigation are: 

\begin{enumerate}

\item The photometric and kinematic analysis shows that NGC 7733 and NGC 7734 have complex structures. The 2-D image decomposition shows that both galaxies contain a disc-like bulge. NGC 7734 also contains a central classical bulge inside a disc-like bulge, thus the possibility of having a composite bulge.

\item A nuclear disc with a semi-major axis of 1.113 kpc has been detected at the centre of NGC 7733, using both photometry and kinematic analysis. The strong anti-correlation between $V/\sigma$ and $h_{3}$ indicates the presence of circular orbits in the centre, confirming the photometric discovery of the nuclear disc.

\item We observed an anti-correlation between $V/\sigma$ and $h_{3}$ at the centre of NGC 7734 but found no specific nuclear structure, however, we did detect the presence of diffuse light. Our study has identified another red and massive component along the bar of NGC 7734, designated as NGC 7734S which is likely a smaller galaxy that traversed the centre of NGC 7734, resulting in dynamically hotter stellar and gas kinematics, as evidenced by the high stellar and gas velocity dispersion observed. The higher dispersion also supports the presence of a central classical bulge in the galaxy.

\item We also found that both galaxies show ongoing SF using $FUV$ and $H\alpha$. NGC 7734 has some recent SF along the bar but NGC 7733 shows bar quenching. As discussed earlier, we did not present the total SFR estimation in this work because our estimate agrees with that of \citet{2023MNRAS.526..198Y}. However, when comparing the SFR along the bar with the total SFR, we found that the SFR in the bar region dominates the overall SFR in NGC 7734. 

\item In NGC 7733, no SFCs were found along the bar. It houses a Seyfert 2 AGN, which may suggest that the AGN feedback (Inside-out feedback) has impacted the central (kpc scale) SF, leading to the suppression of bar-related SF in the galaxy while not affecting the overall SF in the galaxy.

\end{enumerate}

The above findings present an interesting case of an interacting system involving two barred spiral galaxies, each hosting an AGN. Notably, one of the galaxies exhibits enhanced SF along its bar, while the other shows an absence of such activity. Although numerous studies have investigated the connection between stellar bar, SF and AGN, most of these have been done for isolated galaxies.   Therefore, studying these interacting systems provides valuable insights into the relationship between SF along the bar and AGN activity and the broader evolutionary processes at play within these systems. It underscores the importance of conducting a comprehensive statistical analysis of a larger galaxy sample to understand these complex phenomena better. 


\section*{Acknowledgements}
We thank the anonymous referee for the thoughtful review and valuable suggestion, which improved the impact and clarity of this manuscript. We want to thank the anonymous referee for the valuable comments that improved the quality of this work. MD and SB acknowledge the support of the Science and Engineering Research Board (SERB) Core Research Grant CRG/2022/004531 for this research. SK thanks Prof. Arif Babul for the helpful discussions. This paper has used the observations collected at the European Southern Observatory under ESO program 0103.A-0637. This publication uses data from the Indian Space Research Organisation's (ISRO) AstroSat mission, housed at the Indian Space Science Data Centre (ISSDC). The UVIT data used was processed at IIA by the Payload Operations Centre. The UVIT was developed in collaboration with IIA, IUCAA, TIFR, ISRO, and CSA. This research has also used NIR data from IRSF at the South African Astronomical Observatory (SAAO). This research has made use of NASA/IPAC Extragalactic Database (NED), which is operated by the Jet Propulsion Laboratory, California Institute of Technology (Caltech) under contract with NASA. This research has made use of data obtained with the Dark Energy Camera (DECam), which was constructed by the Dark Energy Survey (DES) collaboration. This research is based on observations made with the NASA/ESA Hubble Space Telescope obtained from the Space Telescope Science Institute, which is operated by the Association of Universities for Research in Astronomy, Inc., under NASA contract NAS 5–26555. These observations are associated with proposal ID 15446.  


\section*{Data Availability}
1) UVIT L1 data can be downloaded from \url{https://astrobrowse.issdc.gov.in/astro_archive/archive/Home.jsp}.\\
2) Optical co-added images can be accessed from \url{https://portal.nersc.gov/cfs/cosmo/data/legacysurvey/dr10/south/coadd/355/3554m660/}.\\
3) Optical IFU data cube can be downloaded from \url{http://archive.eso.org/scienceportal/home}.\\
4) NIR data can be shared on request.


\bibliographystyle{aa}
\bibliography{Reference}


\appendix
\section{Star Formation in different regions of NGC 7734} \label{app:Appendix_A}
As previously discussed in Sec.~\ref{sec:Sec_4.1}, the SF within the galaxy group NGC 7733-34 has been analysed by \citet{Yadav23}. They have identified SFCs in the $FUV$ and $H\alpha$ images of NGC 7733 and NGC 7734, estimating the SFR and $\Sigma_{SFR}$ for these emissions in both galaxies. Notably, no SFCs are observed along the bar of NGC 7733. In contrast, recent SF is detected along the bar of NGC 7734 in both the $FUV$ and $H\alpha$ images. 

\begin{figure*}
	\includegraphics[width=\textwidth]{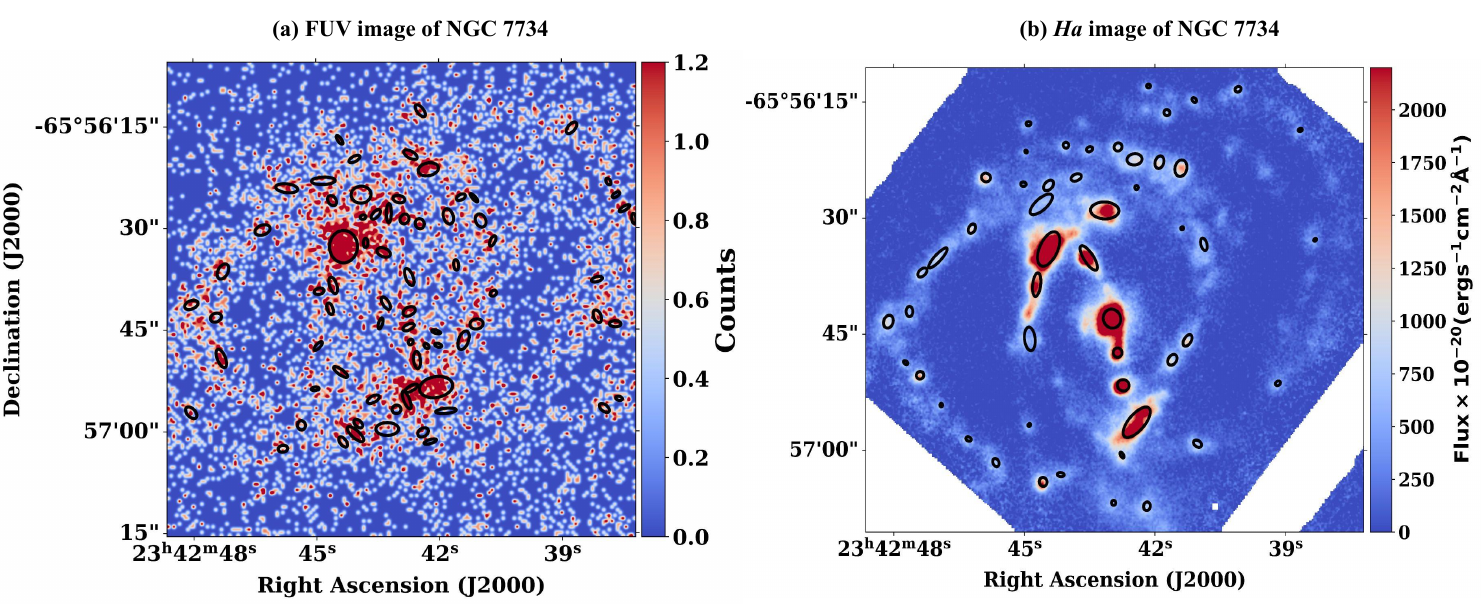}
    \caption{\textbf{Left}: $FUV$ image of NGC 7734 from UVIT, \textbf{Right}: $H\alpha$ narrow band image of NGC 7734 from MUSE. The regions marked in black are the star-forming regions identified using SExtractor. North is up, and east is to the left.}
    \label{fig:Fig_A1}
\end{figure*}

\begin{figure*}
\centering
    \includegraphics[width=0.8\textwidth]{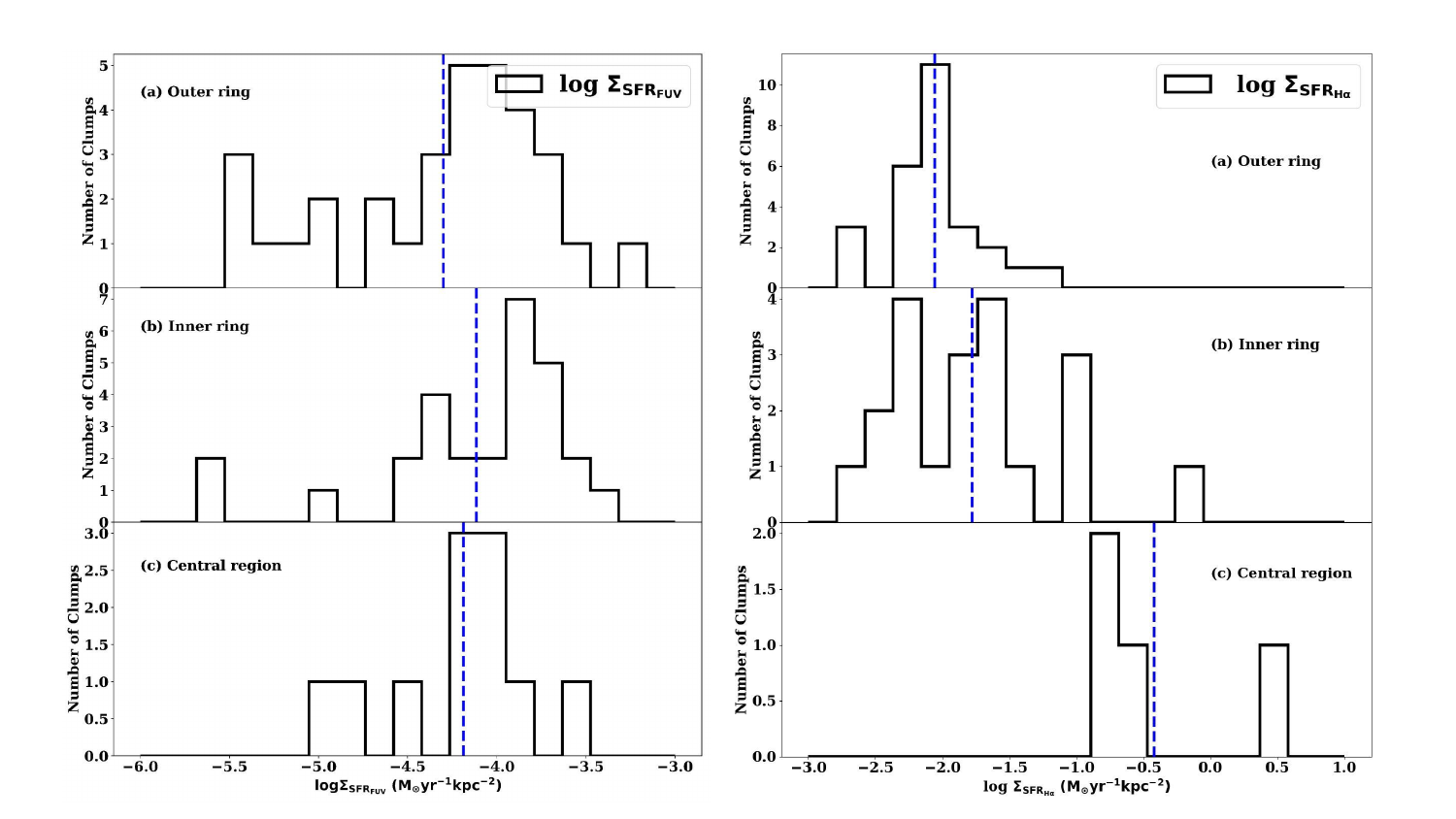}
    \caption{Histogram plot of SFR density in logarithmic scale (log $\Sigma_{SFR}$) of $FUV$ (left) and $H\alpha$ (right) emission in different regions of NGC 7734. \textbf{Left panel}: (a) log $\Sigma_{SFR}$ for outer ring, (b) log $\Sigma_{SFR}$ for inner ring and (c) log $\Sigma_{SFR}$ for central region/bar for $FUV$ emission. \textbf{Right panel}: (a) log $\Sigma_{SFR}$ for outer ring, (b) log $\Sigma_{SFR}$ for inner ring and (c) log $\Sigma_{SFR}$ for central region/bar for $H\alpha$ emission. The blue dashed line represents the median value of log $\Sigma_{SFR}$ in all plots and is tabulated in Table~\ref{tab:Table_A1}.}
    \label{fig:Fig_A2}
\end{figure*}

The identified SFCs in $FUV$ (left) and $H\alpha$ (right) image of NGC 7734 are shown in Fig.~\ref{fig:Fig_A1}. We have divided the galaxy into three sub-regions as discussed in Sec.~\ref{sec:Sec_4.1}. Within these regions, we have estimated the extinction-corrected SFR in $FUV$ and $H\alpha$ using the relations given in Sec.~\ref{sec:Sec_3.5}. We divided each clump's SFR by its area to get an estimate of $\Sigma_{SFR}$. The number of clumps in each region along with the $\Sigma_{SFR}$ is tabulated in Table~\ref{tab:Table_A1}. We compared $\Sigma_{SFR}$ for $FUV$ and $H\alpha$ emission for the three sub-regional clumps. In Fig.~\ref{fig:Fig_A2}, this comparison is shown. 

The left panel of Fig.~\ref{fig:Fig_A2} presents the histogram plot of log $\Sigma_{SFR}$ for $FUV$ emission. It illustrates the log $\Sigma_{SFR}$ distribution for the outer ring, inner ring, and central areas, respectively, from top to bottom. Our analysis indicates a similarity between the distribution of SFCs and the median values of log $\Sigma_{SFR}$ (refer to Table~\ref{tab:Table_A1}), despite there being fewer clumps in the central region compared to the inner and outer rings. Moreover, we observed increased $FUV$ emission at the bar end, leading to high SF. Consequently, the inner ring exhibited a slightly elevated $\Sigma_{SFR}$ compared to the outer ring and central region. 

The right panel of Fig.~\ref{fig:Fig_A2} showcases the histogram plot of log $\Sigma_{SFR}$ for $H\alpha$ emission, presenting the log $\Sigma_{SFR}$ distribution for the outer ring, inner ring, and central region from top to bottom, respectively. It's noteworthy that the central region has a notably high SF compared to the inner and outer rings. Given that $H\alpha$ emission traces the SF of around 10 Myr, there is recent SF activity in the central region of NGC 7734, consistent with the presence of $FUV$ emission. 

The enhanced SF observed along the bar of NGC 7734 may be attributable to the current interaction with NGC 7733 or a past collision with NGC 7734S. It is well-established that galaxy mergers facilitate gas inflows towards the central regions, thereby initiating starburst activity \citep{DiMatteo07, Hopkins08}. Additionally, the non-axisymmetric gravitational potential of the bar plays a crucial role in channelling gas towards the central region, consequently enabling SF. 

\begin{table*}
        \centering
        \caption{SFCs identified in different regions of $FUV$ and $H\alpha$ image of NGC 7734 and the corresponding median value of log $\Sigma_{SFR}$.}
	\begin{tabular}{lcccccr} 
		\hline
            \\
		  & & $FUV$ & & &  $H\alpha$ \\
            
             Region & No. of clumps& & log $\Sigma_{SFR}$ & No. of clumps & &  log $\Sigma_{SFR}$\\
            & & & ($M_{\odot} yr^{-1} kpc ^{-2}$) & & & ($M_{\odot} yr^{-1} kpc ^{-2}$) \\\
            \\
            \hline
            \\
            Outer ring & 32 & & - 4.2998 $\pm$ 0.7714 & 27 & & - 2.0922 $\pm$ 0.0114 \\
            Inner ring & 30 & & - 4.1143 $\pm$ 0.6394 & 20 & & - 1.8678 $\pm$ 0.0058 \\
            Central region/ bar & 11 & & - 4.1867 $\pm$ 0.8453 & 4 & & -0.6752 $\pm$ 0.0007 \\  
            \\
            \hline
	\end{tabular}
	\label{tab:Table_A1}
\end{table*}

\end{document}